\documentclass[aps, prb,twocolumn, superscriptaddress, showpacs,preprintnumbers,amsmath,amssymb]{revtex4-1}

\usepackage{graphicx,color}
\usepackage{multirow}

\begin{document}

\preprint{}

\title{Real-space imaging of quantum Hall effect edge strips}

\author{M.E. Suddards}
\affiliation{
School of Physics and Astronomy,\\}

\author{A. Baumgartner}
\email{andreas.baumgartner@unibas.ch}
\altaffiliation{present address: Nanoelectronics Group, University of Basel, Switzerland}
\affiliation{
School of Physics and Astronomy,\\}

\author{M. Henini}
\affiliation{
School of Physics and Astronomy,\\}
\affiliation{
Nottingham Nanotechnology and Nanoscience Centre, University of Nottingham, Nottingham NG7 2RD, UK\\}

\author{C.J. Mellor}
\affiliation{
School of Physics and Astronomy,\\}

\date{\today}

\begin{abstract}
We use dynamic scanning capacitance microscopy (DSCM) to image compressible and incompressible strips at the edge of a Hall bar in a two-dimensional electron gas (2DEG) in the quantum Hall effect (QHE) regime. This method gives access to the complex local conductance, $G_{\rm ts}$, between a sharp metallic tip scanned across the sample surface and ground, comprising the complex sample conductance. Near integer filling factors we observe  a bright stripe along the sample edge in the imaginary part of $G_{\rm ts}$. The simultaneously recorded real part exhibits a sharp peak at the boundary between the sample interior and the stripe observed in the imaginary part. The features are periodic in the inverse magnetic field and consistent with compressible and incompressible strips forming at the sample edge. For currents larger than the critical current of the QHE break-down the stripes vanish sharply and a homogeneous signal is recovered, similar to zero magnetic field. Our experiments directly illustrate the formation and a variety of properties of the conceptually important QHE edge states at the physical edge of a 2DEG.
\end{abstract}

% - QHE: Novel experimental methods; measurements 
% - Scanning probe microscopes and components
% - Delocalization processes
% - Electron states and collective excitations in multilayers, quantum wells, mesoscopic, and nanoscale systems

\maketitle

% -------- Introduction ----------------------------------------
\section{INTRODUCTION}

The quantum Hall effect\cite{Klitzing_Dorda_Pepper_PRL45_1980} (QHE) is a macroscopic quantum phenomenon in a two-dimensional electron gas (2DEG) where the formation of discrete Landau levels in a large perpendicular magnetic field, $B$, leads to quantized values of the Hall resistance. Although a theoretical understanding was developed shortly after its discovery,\cite{Prange_book} the experimental investigation of the microscopic origin using local probe techniques only became feasible in recent years. The challenges of such experiments are the required cryogenic temperatures, high magnetic fields and that most 2DEGs are buried below a dielectric layer.

The most intuitive picture of the QHE is that of edge states.\cite{Buettiker_PRL57_1986, Haug_SemicondSciTechnol8_1993} If an integer number of Landau levels (LLs) are completely filled (integer filling factor $\nu$) in the bulk of the sample, the density of states, $D$, at the Fermi energy is zero except near the sample edge, where the energy of the LLs is increased by the confining potential. Where the LL energy equals the Fermi level, areas of finite $D$ are formed along equipotential lines. The self-consistent spatial rearrangement of these states (edge reconstruction) leads to the formation of compressible strips, separated by incompressible strips at the sample edge.\cite{Chklovskii_PRB46_1992}

Edge states are crucial for the formation of the quantum Hall plateaus by separating the states at opposite edges.\cite{Siddiki_Gerhardts_PRB70_2004} Delocalized low-energy excitations are possible only in the compressible regions, which allows, for example, electrically controlled interference experiments with electrons.\cite{Ji_Heiblum_Nature422_2003, Neder_Heiblum_Nature448_2007, Bieri_PRB79_2009} Edge states are also relevant for topologically protected states in the fractional QHE,\cite{Nayak_Das_Sarma_RevModPhys80_2008} or in novel topological states at zero magnetic field.\cite{Koenig_Science318_2007, Maciejko_annurev_conmatphys}

Edge states localized at potential fluctuations in the bulk of a macroscopic 2DEG have been imaged using various scanning probe techniques.\cite{Tessmer_Ashoori_Nature392_1998, Yacoby_SolidStateComm111_1999, Ilani_Yacoby_Nature427_2004, Hashimoto_Morgenstern_PRL101_2008} Scanning gate measurements provide the link between these microscopic results and actual transport experiments in real devices, e.g. by demonstrating an enhanced sensitivity of the device resistances to local variations of the potential landscape.\cite{Kicin_Pioda_Ihn_PRB70_2004, Baumgartner_PRB76_2007, Hackens_NatComm1_2010} The position of the innermost incompressible strip and the question of where the excess current is carried was investigated by locally measuring the electrostatic force gradient due to the Hall potential over a biased sample.\cite{Ahlswede_Weitz_PhysicaB298_2001}

Here we present dynamical scanning capacitance microscopy\cite{Baumgartner_RSI80_2009} (DSCM) images recorded at the edge of a 2DEG in the QHE regime at zero bias as well as in the QHE breakdown regime at finite currents. A related method was used recently to image square areas of a 2DEG at $\nu\approx2$.\cite{Lai_PRL107_2011} The DSCM signals are directly related to the complex local sample conductance, which comprises the local quantum capacitance and the local conductivity of the 2DEG. Our instrument has a large scan area and a relatively high scan rate which allows to image edge states over large distances in two dimensions, leading to very clear and intuitive illustrations of edge states over several QHE periods.

In section II we describe the working principle of a DSCM, while we present DSCM experiments on a 2DEG at zero magnetic field in section III and and in the QHE regime in section IV. These results are discussed in section V and the QHE breakdown is investigated in section VI. Our results are summarized in section VII.

\section{Dynamical Scanning Capacitance Microscopy}

We use a home-built dynamic scanning capacitance microscope\cite{Baumgartner_RSI80_2009} (DSCM) which operates between $1.9\,$K and room temperature and magnetic fields up to $12\,$T. The microscope is based on a non-contact atomic force microscope (AFM) in a low pressure ($2\,$mbar) He atmosphere. A quartz tuning fork sensor and a phase-locked loop allow non-optical sensor excitation and readout of the topography and the average force. An amplitude feedback loop maintains a constant oscillation amplitude and allows us to measure the mechanical sensor dissipation.\cite{Rychen_Ihn_Enssin_RSI71_2000} The AFM tip is fabricated from an electrochemically sharpened PtIr wire and an electrical connection enables simultaneous AFM/DSCM measurements.

\begin{figure}[t]{
\centering
\includegraphics{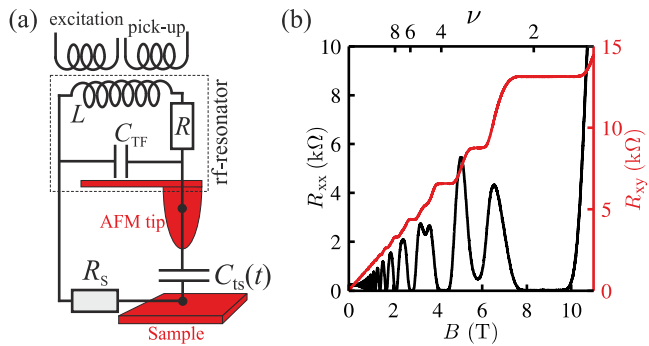}
}
\caption{(a) Schematic of the detection circuit showing the rf resonator, the excitation and pick-up coils and the tip-sample capacitance in series with the sample conductance coupled to the resonator. (b) Magnetotransport data of the Hall bar at $T=1.9\,$K.}
\end{figure}

The complex tip-sample conductance is measured using a radio frequency (rf) resonator. A small hand-made copper coil $L$ is connected to the AFM tip and forms a resonant circuit with the stray capacitance of the tuning fork $C_{\rm TF}$ and the lead resistances $R$, see Fig.~1a. The transmission of the resonator is perturbed by the tip-sample capacitance $C_{\rm ts}$, which is probed by an rf excitation and a pick-up coil. We use high-frequency lock-in amplifiers to measure the resulting voltage on the pick-up coil, $U_{\rm out}$,  at the resonance frequency of the unperturbed resonator ($\sim130\,$MHz). An analysis of the tip-sample interaction in terms of lumped-circuit elements is possible since the wavelength of the excitation is much larger than the relevant geometrical dimensions. One can show\cite{Baumgartner_RSI80_2009} that for reasonably sharp tips $U_{\rm out}$ is proportional to the complex conductance between the tip and ground, $G_{\rm ts}$, independent of the details of the tip-sample interaction:

\begin{equation}
U_{\rm out}\propto G_{\rm ts}.
\end{equation}

The setup-specific calibrated proportionality constant includes the excitation voltage. For a homogeneous sample, $G_{\rm ts}$ is given by the tip-sample capacitance, $C_{\rm ts}$, in series with the complex sample conductance from the tip position to ground. The voltage amplitude on the AFM tip we estimate to be $24\,$mV, a compromise between a sufficient signal-to-noise ratio and useful scan rates and minimal charging of the sample surface.

The perturbation of the rf resonator due to the sample is exceedingly small and a direct measurement is very slow and not suitable for scanning.\cite{Baumgartner_RSI80_2009} We operate the AFM in the dynamic mode where the tip oscillates perpendicular to the sample surface. The tip motion modulates the tip-sample capacitance and thus the extracted conductance with an amplitude $G_{\rm ts, \omega}$ at the frequency $\omega$ of the AFM sensor (and its harmonics). This signal is recorded simultaneously with the AFM data using low-frequency lock-in amplifiers.
 
The DSCM technique allows us to measure output voltage differences of $\sim1\,$nV at the required scan speeds, which corresponds to a capacitance resolution of $\sim0.5\,$aF. Furthermore, DSCM does not require calibration or subtraction of base-line data. The lateral resolution of $\sim100\,$nm is mainly limited by the point of closest approach to the surface, where the change of the tip-sample capacitance is the largest. This resolution is comparable to other scanning capacitance experiments with the tip in contact to the sample, but without the detrimental mechanical abrasion of the tip apex. Features in the DSCM images with a seemingly better lateral resolution can be traced back to changes in the average tip-sample distance due to topographic features on the surface, which are resolved with typical AFM resolution.\cite{Suddards_Baumgartner_PhysicaE40_2008}

DSCM signals can be understood intuitively by considering a purely ohmic sample resistor $R$ in series with the tip-sample capacitance, $C_{\rm ts}$, as shown in Fig.~1a. For highly conductive samples, i.e. $\omega_{\rm rf}RC_{\rm ts}<<1$, one finds

\begin{equation}
G_{\rm ts, \omega}\approx \frac{dG_{\rm ts}}{dC_{\rm ts}}C_{\rm ts, \omega} = \omega_{\rm rf} C_{\rm ts, \omega} (2\omega RC_{\rm ts} +i)
\end{equation}
with the capacitance variation $C_{\rm ts, \omega}$ at the frequency $\omega$ of the tuning fork oscillation. In this limit the imaginary part of the signal is given by the capacitance variation over a tuning fork oscillation cycle. For large sensor oscillation amplitudes this reflects directly the local tip-sample capacitance. The real part is given by the $RC$-constant of the tip-sample capacitance and the sample resistance to ground and thus contains information on the local sample resistivity and dissipation. For $R\rightarrow 0$ the real part is zero. In the limit of $\omega_{\rm rf}RC_{\rm ts}>>1$ and $R\rightarrow \infty$ both, the real and and the imaginary parts become zero because the tip-sample capacitance can not be charged during an rf cycle. In the QHE regime, both limits are conceivable since the 2DEG can contain locally well conducting and poorly conducting regions and the 2DEG's quantum capacitance can go to zero with diminishing density of states.

\section{2DEG at zero field}

We performed DSCM measurements at a temperature of $T=2.1\,$K at the edge of a $500\,\mu$m wide and $5\,$mm long Hall bar defined by wet-chemical etching in a remotely doped AlGaAs/GaAs heterostructure with a 2DEG $50\,$nm below the sample surface.\cite{Shashkin_Kent_Eaves_PRB49_1994} The magneto-transport characteristics are shown in Fig.~1b and exhibit well-developed quantum Hall plateaus in the transverse resistance, $R_{\rm xy}$, and Shubnikov -de Haas (SdH) oscillations in the longitudinal resistance, $R_{\rm xx}$. The Hall slope and the SdH oscillations both yield an electron density of $\sim3.9\times 10^{15}\,$m$^{-2}$ and we find a zero-field Drude mobility of $\sim45\,$m$^2$/Vs. Henceforth, we use the term `filling factor' with reference to these bulk properties. Unless stated otherwise, the DSCM experiments were performed without passing a current through the sample.

\begin{figure}[t]{
\centering
\includegraphics{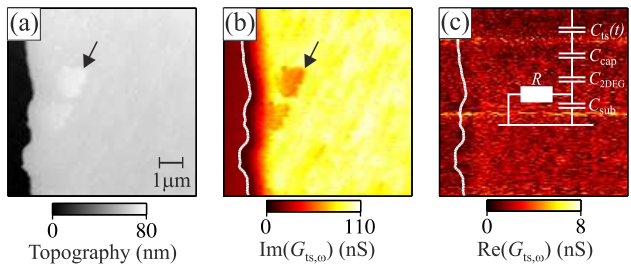}
}
\caption{(a) Topography of the sample, (b) imaginary and (c) real part of the tip-sample conductance $G_{\rm ts}$ at zero magnetic field as a function of the sensor position. The white contour lines indicate the sample edge extracted from (a) with the 2DEG to the right.}
\end{figure}

Figures~2(a), (b) and (c) show the topography of the sample and the simultaneously recorded imaginary and real part of $G_{\rm ts,\omega}$, respectively, in a scan across the Hall bar edge at $B=0$. The white contour lines overlaid on all images are extracted from the topography and indicate the position of the physical edge of the Hall bar. The real part in Fig.~2(c) is essentially zero and shows no structure. In contrast, the magnitude of the imaginary part in Fig.~2(b) is significantly larger on the Hall bar than over the etched region. On the 2DEG the signal is homogeneous except for small variations introduced by the topography, as for example highlighted by arrows in Figs.~2(a) and 2(b).\cite{Suddards_Baumgartner_PhysicaE40_2008} The physical edge does not correspond exactly to the onset of the capacitance signal, which we attribute to a depletion of the 2DEG near the edge due to missing dopants and the increased surface potential leading to the 2DEG confinement. We note a small minimum close to the edge of the Hall bar where the signal is reduced below the value far from the 2DEG. A simple finite-element method (FEM) simulation confirms that this is a pure geometrical effect whereby the edge step reduces the capacitive coupling between the tip and sample.\cite{Matt_thesis}

The amplitude of the imaginary part of the signal can be interpreted using the lumped circuit model in the inset of Fig.~2(c).\cite{Tessmer_Ashoori_PRB66_2002} The coupling between the tip and the 2DEG can be described by three capacitances in series; the tip-surface capacitance, $C_{\rm ts}(t)$; the capacitance of the GaAs cap layer, $C_{\rm cap}$, and the self or quantum capacitance $C_{\rm 2DEG}$ of the 2DEG. The current path to ground is closed by the capacitance of the substrate, $C_{\rm sub}$, in parallel with the resistance through the 2DEG, $R$. We estimate $R$ from the homogeneous zero-field bulk conductivity and a circular-symmetric current density to the tip.\cite{Baumgartner_RSI80_2009, Matt_thesis} Simple modeling indicates that except for very large $R$ the substrate capacitance can be neglected. Therefore, the impedance $G_{\rm ts}$ can be reduced to the tip-sample capacitance in series with the sample resistance.

\section{Quantum Hall effect regime}

\begin{figure}[b]{
\centering
\includegraphics{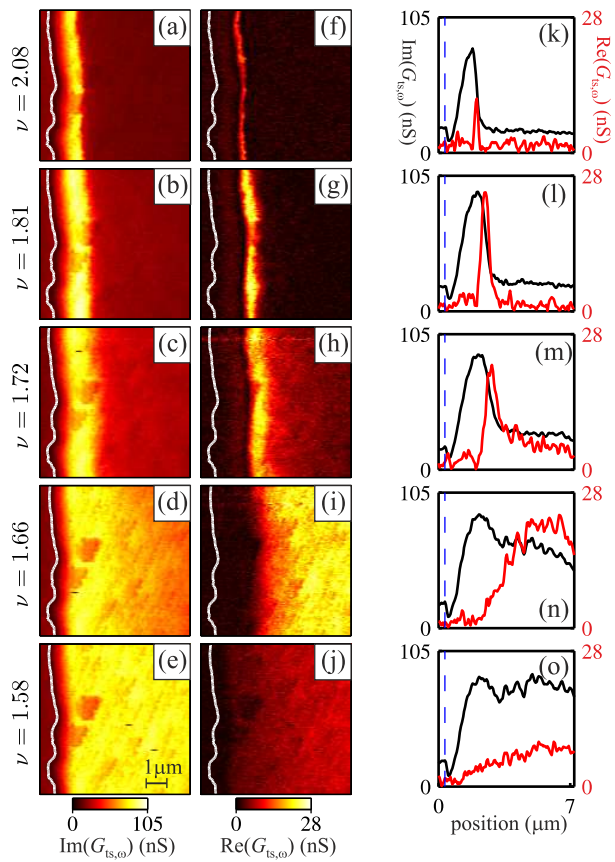}
}
\caption{(a)-(e) imaginary and (f)-(j) real part of $G_{\rm ts,\omega}$ at filling factors between $\nu=1.58$ and $\nu=2.08$ at the sample edge. (k)-(o) cross sections perpendicular to the sample edge.}
\end{figure}

Figure~3 shows DSCM scans in the QHE regime between bulk filling factor $\nu=2.08$ and $\nu=1.58$. The left-hand column, images (a)-(e), shows the imaginary part of $G_{\rm ts,\omega}$; the central column, images (f)-(j), shows the real part of $G_{\rm ts,\omega}$. Cross-sections perpendicular to the sample edge are plotted in the right-hand column, figures (k)-(o).

At $\nu=2.08$ the imaginary part of the DSCM signal shows a $\sim1\,\mu$m wide bright stripe close to the sample edge. The signal level in the interior of the sample is the same as the background over the etched region. Both findings are in stark contrast to the homogeneous signal over the 2DEG observed at zero magnetic field. The maximum signal on the stripe, however, is comparable to the signal at zero magnetic field.
The real part of the DSCM signal shows a narrow maximum at the boundary between the stripe in the imaginary part and the sample interior. It occurs where the imaginary part decays to approximately half of its maximum value, and it has a width that is comparable to the distance over which the imaginary part drops to the background level in the sample interior. In the interior of the sample the real part is negligible, similar to that at zero magnetic field.

\begin{figure}[t]{
\centering
\includegraphics{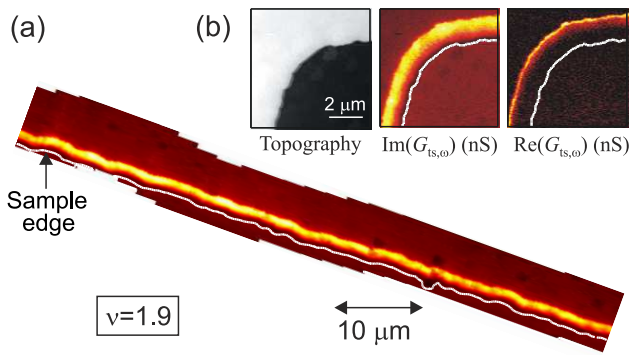}
}
\caption{(a) Composite image of several scans of Im$(G_{\rm ts,\omega})$ over a large distance along the edge of the Hall bar. (b) Topography, Im$(G_{\rm ts,\omega})$ and Re$(G_{\rm ts,\omega})$ at a corner of a Hall bar contact.}
\end{figure}

\begin{figure}[b]{
\centering
\includegraphics{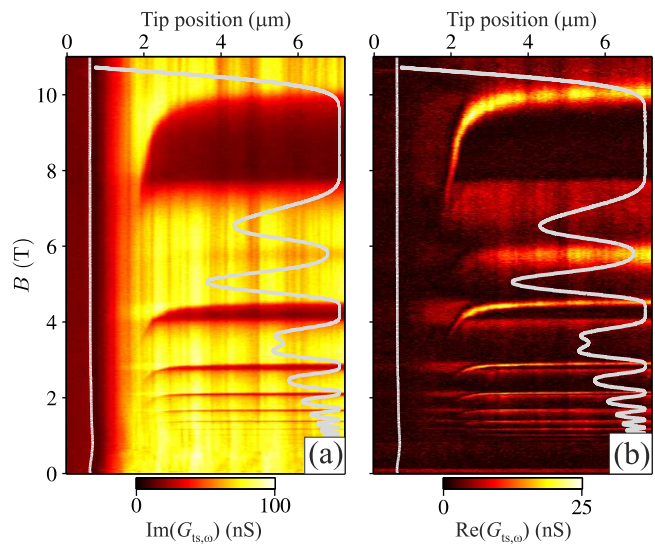}
}
\caption{DSCM line scans at the sample edge of (a) Im$(G_{\rm ts,\omega})$ and (b) Re$(G_{\rm ts,\omega})$ as a function magnetic field. $R_{\rm xx}$ is overlayed in both images for orientation.}
\end{figure}

The features at $\nu=1.81$ are similar in form to those at $\nu=2.08$. However, the width of the stripe in the imaginary part has increased by approximately $50\%$ compared to $\nu=2.08$ and the signal does not return exactly to the background level in the sample interior. At $\nu=1.72$ the stripe structures remain clearly defined in both the real and imaginary part. However, both show an increased background in the interior of the sample. This background increases considerably for $\nu=1.66$, where the stripe structures are difficult to distinguish and both signals are relatively large in the bulk. We note that the stripe maxima in the imaginary part have not changed significantly between the different filling factors. At $\nu=1.58$, near a maximum of $R_{\rm xx}$, the imaginary part image is almost identical to $B=0$, see Fig.~2, e.g. the response is homogeneous across the sample. The real part, though smaller than at $\nu=1.66$, still exhibits a shallow monotonic increase away from the edge.

Fig.~4a shows a composite image of numerous scans of Im$(G_{\rm ts,\omega})$ along $\sim100\,\mu$m of the sample edge at $\nu=1.90$. In previous experiments we found that the stripe width depends on the edge confinement potential.\cite{Suddards_Baumgartner_PhysicaE40_2008} Here we observe no significant variation in the shape or the amplitude of the stripes. This also holds for bent edges at Hall bar contacts, as illustrated in Fig.~4b. These measurements suggest that the electron density and the edge confinement do not vary significantly even over macroscopic distances in this sample.

The described structures in the DSCM images are reproduced $1/B$-periodically at lower magnetic fields. As an illustration, Fig.~5 shows line scans across the edge as a function of magnetic field at a fixed position along the sample. The periodicity is consistent with the electron density extracted from the transport data. For comparison, the longitudinal resistance of the Hall bar is plotted in grey on the same field axis.

Considering the imaginary part of the signal in Fig.~5(a) we find that at magnetic fields where $R_{\rm xx}$ is zero the stripe structure is well defined and the signal in the bulk is at a minimum. Away from $R_{\rm xx} = 0$ the signal over the 2DEG becomes more homogeneous until, at fields between integer filling factors where $R_{\rm xx}$ has a maximum, the scans become comparable to the ones at zero field. The peak in Re$(G_{\rm ts,\omega})$ follows the inner boundary of the stripe in Im$(G_{\rm ts,\omega})$ and diverges into the sample interior at the high-field ends of the $R_{\rm xx} = 0$ intervals, see Fig.~5(b). At these fields the width of the Re$(G_{\rm ts,\omega})$ stripe, measured perpendicular to the sample edge, increases continuously. At the low-field ends, the bulk signal of Re$(G_{\rm ts,\omega})$ is clearly different from zero, but the stripe is still visible. Around $\nu=3$, where the spin-degeneracy of the Landau levels is lifted by the Zeeman energy, the imaginary part is slightly suppressed in the bulk and a bright stripe in the real part occurs.

\section{Interpretation and Discussion}
We consider the lumped circuit representation of the experiment shown in Fig.~2c. Neglecting the substrate capacitance, the total complex conductance is given by the tip-surface and the 2DEG capacitance in series with the resistance of the 2DEG from the tip-position to ground and can be described by Eq.~2 for small enough sample conductances. The capacitance variation due to the tuning fork oscillation is approximately constant throughout a scan, so that the contrast in our images, apart from topographic and geometric artefacts, is determined solely by the local properties of the 2DEG.

In this simplified picture variations in Im$(G_{\rm ts,\omega})$ depend on the quantum capacitance of the 2DEG, which is proportional to the density of states at the Fermi energy $D(E_{\rm F})$. Variations in the real part, Re$(G_{\rm ts,\omega})$, are determined by the effective $RC$-charging time of the 2DEG at the sensor position. If the AFM-cycle average of $C_{\rm ts}$ remained constant over a scan, the variations of Re$(G_{\rm ts,\omega})$ could simply be interpreted as changes in the local sample conductivity. However, in the QHE regime both, $R$ and the 2DEG's capacitance vary with position and their contributions can not be disentangled without model assumptions.\cite{Matt_thesis} Additional complexity arises from the fact that the AFM tip couples to a relatively large area so that the finite size of the tip and the inhomogeneity of the sample characteristics should be considered.

Based on the above model the bright stripe in Im$(G_{\rm ts,\omega})$ can be interpreted as the compressible regions at the edge of the sample. The model assumptions are valid over these regions as $D(E_{\rm F})$ is large and the resistance to ground relatively small. The signal strength is similar to the zero-field scans, which adds weight to the conclusion that the signal is determined by the capacitive part, while the resistive part is negligible. Theoretical models suggest that the compressible regions at the edge are separated from the sample interior by a large resistance across the innermost incompressible strip.\cite{Tessmer_Ashoori_PRB66_2002, Siddiki_Gerhardts_PRB70_2004} In our experiments, Re$(G_{\rm ts,\omega})$ measures the $RC$-time constant of the local portion of the 2DEG. It increases at the boundary between the compressible strip and the sample interior and is consistent with the inner-most incompressible strip insulating the sample interior from the contacts. In the bulk of the sample the signal is further suppressed due to $D(E_{\rm F})$ tending to zero. Our results support the general expectation that around integer filling factors the bulk is incompressible, i.e. $D(E_{\rm F})\approx 0$. Between integer filling factors the incompressible strip moves into the sample interior and the whole 2DEG becomes compressible. At these fields the coupling between edge states increases, which leads to the transition between the quantized QHE plateaus.\cite{Baumgartner_PRB76_2007} The simultaneous increase in back-scattering leads to the finite longitudinal resistance. Our experiments therefore illustrate how the strip position determines the transport characteristics at a given magnetic field.

The compressible and incompressible strips develop due to the non-linear screening of the edge potential by the 2DEG in the QHE regime.\cite{Chklovskii_PRB46_1992} In contrast to early theoretical models, our experiments show that the incompressible strips do not diverge into the bulk at integer bulk filling factors. This was also found in more recent numerical calculations\cite{Siddiki_Gerhardts_PRB70_2004} and attributed to the absence of long-range potential fluctuations in the model. Such fluctuations might not be dominant in our device because of the shallow cap layer and low impurity concentrations.

We do not resolve any signal variations that might indicate the presence of other incompressible stripes with a smaller local filling factor which one might expect closer to the sample edge. This is in agreement with theoretical results\cite{Siddiki_Gerhardts_PRB70_2004} suggesting that these strips are very narrow (i.e. below the resolution of the DSCM) and that they do not insulate the compressible states from the contacts. In our experiment we observe that when the innermost incompressible strip becomes very narrow the real part of the DSCM signal in the bulk of the sample is enhanced. This suggests that the resistance of the innermost compressible strip decreases with its width.

\section{Breakdown of the QHE}

\begin{figure}[b]{
\centering
\includegraphics{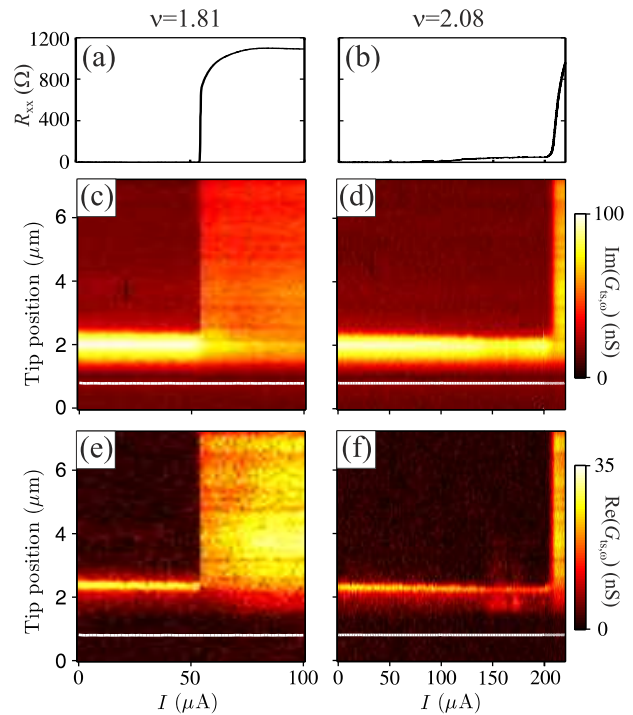}
}
\caption{(a) and (b) Longitudinal resistance of the sample vs. the total current $I$ imposed on the Hall bar for filling factors $\nu=1.81$ and $\nu=2.08$, respectively. (c)-(f) DSCM line scans at the sample edge of the imaginary [(c) and (d)] and the real part [(e) and (f)] of $G_{\rm ts,\omega}$ as a function of $I$ (the 2DEG resides to the top of the white line).}
\end{figure}

A poorly understood phenomenon is the breakdown of the QHE at large applied currents.\cite{Shashkin_Kent_Eaves_PRB49_1994, Nachtwei_PhysicsE4_1999} In Figs.~5(a) and (b) the longitudinal resistance $R_{\rm xx}$ is plotted as a function of the applied total current $I$ for the filling factors $\nu=1.81$ and $\nu=2.08$ [the current is given on the axis of Figs.~5(e) and (f)].
At $\nu=1.81$ the longitudinal resistance increases sharply above a critical current $I_{\rm c}\approx50\,\mu$A, from zero to $R_{\rm xx}\approx 1\,$k$\Omega$, where it saturates. This resistance is considerably larger than the zero magnetic field value. At $\nu=2.08$ we observe in Fig.~5(b) a weak continuous increase of the resistance starting around $I=50\,\mu$A to $R_{\rm xx}\approx 100\,\Omega$, followed by a sharp increase to $R_{\rm xx}\approx 1\,$k$\Omega$ at $I_{\rm c}\approx200\,\mu$A.

Figures~5(c) to (f) show DSCM line scans as a function of the current $I$ at the two filling factors discussed above; (c) and (d) show the imaginary, (e) and (f) the real part of the signal. For all currents smaller than $I_{\rm c}$ both signals exhibit the same stripe structure as at zero current. At $I_{\rm c}$ the pattern changes abruptly from the narrow stripe at the edge of the sample to a much more homogeneous signal across the 2DEG. In the imaginary part at $\nu=1.81$, shown in Fig.~5(c), the bright stripe does not end abruptly at $I_{\rm c}$, but decreases in width when further increasing the current. In contrast, the bulk of the sample switches abruptly from zero to a finite value, which corresponds approximately to half of the zero-field and zero-bias experiment. The stripe observed in the real part immediately disappears at $I_{\rm c}$. The behavior at $\nu=2.08$ is generally similar to $\nu=1.81$, except that the critical current is much larger. However, while we observe a finite longitudinal resistance already below $I_{\rm c}$, the stripe structure in the DSCM images at these intermediate currents does not deviate significantly from the zero-current structure, apart from a slight narrowing and a small decrease in amplitude. The images do not change under reversal of the bias current.

At zero-bias the incompressible strip, mainly observed in the real part of the signal, is slightly wider at $\nu=1.81$ than at $\nu=2.08$. In addition the imaginary part shows a finite signal in the sample interior, see Fig.~3. From the latter one can deduce that the density of states at the Fermi energy in the interior is not zero. We adopt the following picture of the QHE breakdown:\cite{Tsemekhman_Wexler_Thouless_PRB55_1997} at a large enough bias the localized compressible strips in the bulk follow the equipotential lines, which become distorted by the Hall potential for finite currents. At the critical current the localized states percolate and potential fluctuations become irrelevant. This transition from inhomogeneous to homogeneous sample properties happens on a very small current interval, as observed in our experiments. The differences found for different filling factors might then be due to the different insulating and screening properties of the compressible and incompressible strips at a given magnetic field.

% -------- Conclusion and Outlook ---------------------------
\section{CONCLUSION}
We have imaged the physical edge of a two-dimensional electron gas in the quantum Hall effect regime using dynamical scanning capacitance microscopy. We interpret our measurements as spatial images of the local complex tip-sample-ground conductance. We find bright stripes in both, the imaginary and the real part of the signal around integer filling factors, whereas at fields where the longitudinal resistance has a maximum, we find essentially homogeneous signals comparable to the scans at zero magnetic fields. We argue that the stripes are directly related to the compressible and incompressible strips forming at the edge of the sample as a consequence of the non-linear screening of the 2DEG in the QHE regime. We find a surprisingly good match with a theoretical model of edge states and the resulting magnetoresistance. An investigation of these edge strips as a function of the applied current shows an abrupt transition from the strongly inhomogeneous DSCM signals at low currents to essentially homogeneous signals for currents larger than the critical current obtained for the QHE break-down in transport experiments.

\section{ACKNOWLEDGEMENTS}
This work is financially supported by the Engineering and Physical Sciences Research Council (UK).

\bibliographystyle{apsrev}

\end{document}